# Progress Report: A Deep Learning Guided Exploration of Affine Unimodular Loop Transformations

Massinissa Merouani
New York University Abu Dhabi
mm12191@nyu.edu

Khaled Afif Boudaoud
Ecole Nationale Supérieure
d'Informatique
hk_boudaoud@esi.dz

Iheb Nassim Aouadj
Ecole Nationale Supérieure
d'Informatique
hi_aouadj@esi.dz

Nassim Tchoulak
Ecole Nationale Supérieure
d'Informatique

Fatima Benbouzid-Sitayeb
Ecole Nationale Supérieure
d'Informatique

Karima Benatchba
Ecole Nationale Supérieure
d'Informatique

Hugh Leather
Meta AI

Riyadh Baghdadi
New York University Abu Dhabi
baghdadi@nyu.edu

## Abstract

In this paper, we present a work in progress about a deep learning based approach for automatic code optimization in polyhedral compilers. The proposed technique explores combinations of affine and non-affine loop transformations to find the sequence of transformations that minimizes the execution time of a given program. This exploration is guided by a deep learning based cost model that evaluates the speedup that each sequence of transformations would yield. Preliminary results show that the proposed techniques achieve a 2.35x geometric mean speedup over state of the art polyhedral compilers (Pluto).

## 1 Introduction

Automatic code optimization is a long-sought goal in the compiler community, It allows the generation of highly optimized code without requiring extensive development effort or domain expertise. For an automatic code optimization tool to be effective, it needs to be able to quickly find the set of legal transformations that achieve the desired goal (e.g., minimize the execution time of the input program). This task is challenging for two main reasons. First, finding the best sequence of transformations is hard because of the large space of possible transformations. Since exhaustive search is impossible, automatic optimization tools need to use efficient search techniques while exploring this space. The second challenge is the need for fast and accurate cost models used to evaluate the quality of candidate transformations. Evaluating candidates by executing them renders most techniques too slow and impractical [6]. Therefore, using deep learning based cost models for assessing the quality of candidate transformations is an interesting alternative [1, 2, 5].

The Tiramisu auto-scheduler [2] is an automatic code optimization module included in the Tiramisu compiler [3]. It allows exploring sequences of code transformations that include loop fusion, interchange, parallelization, tiling, and unrolling using tree-based search techniques. It relies on a deep learning based cost model for steering the exploration towards finding interesting transformations. The cost model is an LSTM based neural network that takes as input a set of simple features representing the unoptimized code and a sequence of code transformations. This model works by recursively embedding a program depending on its AST (Abstract Syntax tree) structure then, from the final embedding, predicts the performance of the given transformations. Many of the code transformations are simply represented as boolean tags. For example, there is a boolean tag for each loop level indicating whether the loop is parallel. There is another tag indicating whether a given loop was interchanged. While such a simple representation is enough for certain non-affine transformations such as parallelization and vectorization, it is not expressive enough to represent the whole space of affine transformations. For example, it is not well suited to represent a combination of an arbitrary number of affine transformations and cannot capture the order in which these transformations are applied. The goal of this project is to solve this problem. We achieve this by extending the Tiramisu auto-scheduler in two different ways:

- First, extend the cost model: instead of having a cost model that takes a simple list of features as input (e.g., use boolean tags to represent transformations), our goal is to build a new model that takes the polyhedral representation of the code and code transformations as input. Both, code and code transformations will be represented as constraint matrices (iteration domain matrix, schedule matrix, and array access matrices). The new model also takes other non-polyhedral features (that were used in the original model), such as boolean tags for non-affine transformations (parallelization, vectorization ...).

- Second, adapt the search space exploration technique to cover the whole space of unimodular affine transformations. The goal here is to become able to apply an arbitrary combination of unimodular transformations (unlike the original technique which only supported loop interchange).

In this paper, we present our work-in-progress in order to achieve these goals. So far we have achieved the following:

- We have extended the deep learning based cost model in [2] to support unimodular affine transformations. The original model only supported loop interchange. The format of input to the model is now a schedule matrix instead of a tag based representation.
- We have adapted the search space exploration in [2] to cover the whole space of unimodular affine transformations.

## 2 Search Space Exploration

We explore the space in two stages: first, we explore affine unimodular transformations then we explore non-affine transformations (parallelization, vectorization, ...).

In the first stage, in order to cover the whole space of unimodular affine transformations, we explore combinations of loop skewing, reordering, and reversal applied multiple times in any order[8]. Each of these sequences is represented using a single schedule matrix which is the product of the respective schedule matrices of each transformation in the sequence. The objective of this stage of the exploration is to find affine transformation matrices that would improve the data locality and enable thread-level and SIMD-level parallelism.

The second stage of the search space consists in exploring combinations of non-affine loop transformations. The loop transformations that we are considering in this stage are fusion, parallelization, tiling, and loop unrolling.

The exploration of these two stages is performed by a tree-based search algorithm (such as Beam Search and Monte Carlo Tree Search) that will consider the performance of different combinations of these transformations along with their different parameters. Throughout the exploration, illegal transformations (i.e. schedules that violate data dependencies) are detected and pruned (we use classical polyhedral dependence analysis to compute the dependencies).

## 3 Cost Model

Due to the large size of the search space, we need a fast and accurate way to evaluate the transformation candidates that are encountered during exploration. We redesigned the cost model [2] to support schedule matrices along with the previously supported non-affine transformations (non-affine transformations are represented using tags). This cost model takes as input features that represent the program and code transformations, including the schedule matrix, and predicts the expected speedup that would result from applying these transformations to the program. The architecture of the model is an AST (Abstract Syntax Tree) based recursive model and takes an additional vector at its inner layers which represents a learned embedding of the transformation matrix. This embedding vector is generated through a feed-forward layer that summarizes the matrix into a single vector.

For training this model, we used a large corpus of transformed Tiramisu programs for which we applied a large number of different optimization combinations and recorded the speedup of each combination. To create this dataset, we randomly generated a large number of synthetic Tiramisu programs on which we executed the previously described exploration procedure. Then, for each candidate visited during the exploration, we measured its execution time and saved the triplet - program, transformations, and execution time-into our dataset. For a fair evaluation of the model, we make sure that the programs used for evaluation are never encountered during the training even if the applied transformations are different.

## 4 Preliminary Results

As a first iteration of the project and in order to evaluate the potential of the proposed technique, we started by adding two affine transformations to the search space (loop skewing and loop reordering), in addition to having non-affine loop transformations (loop parallelization, loop tiling, and loop unrolling). To explore this space we use beam search, where at each search level we explore different alternatives of a given transformation.

In figure 1, we compare the speedups that our technique achieves against speedups of the transformations found by Pluto [4] using the `--tile --parallel` flags. In this experiment, we evaluate both the model-guided exploration (i.e. where the candidates visited during the beam search exploration are evaluated using the deep learning cost model) and the exploration guided by execution (i.e. where each visited candidate is compiled and executed in order to measure its runtime).

The cost model we used was trained on 14 million randomly generated synthetic data points (each data point is a program with a sequence of transformations). This model was trained using the NDCG-loss2++ [7] loss to sort the different sequences of transformations of a given program

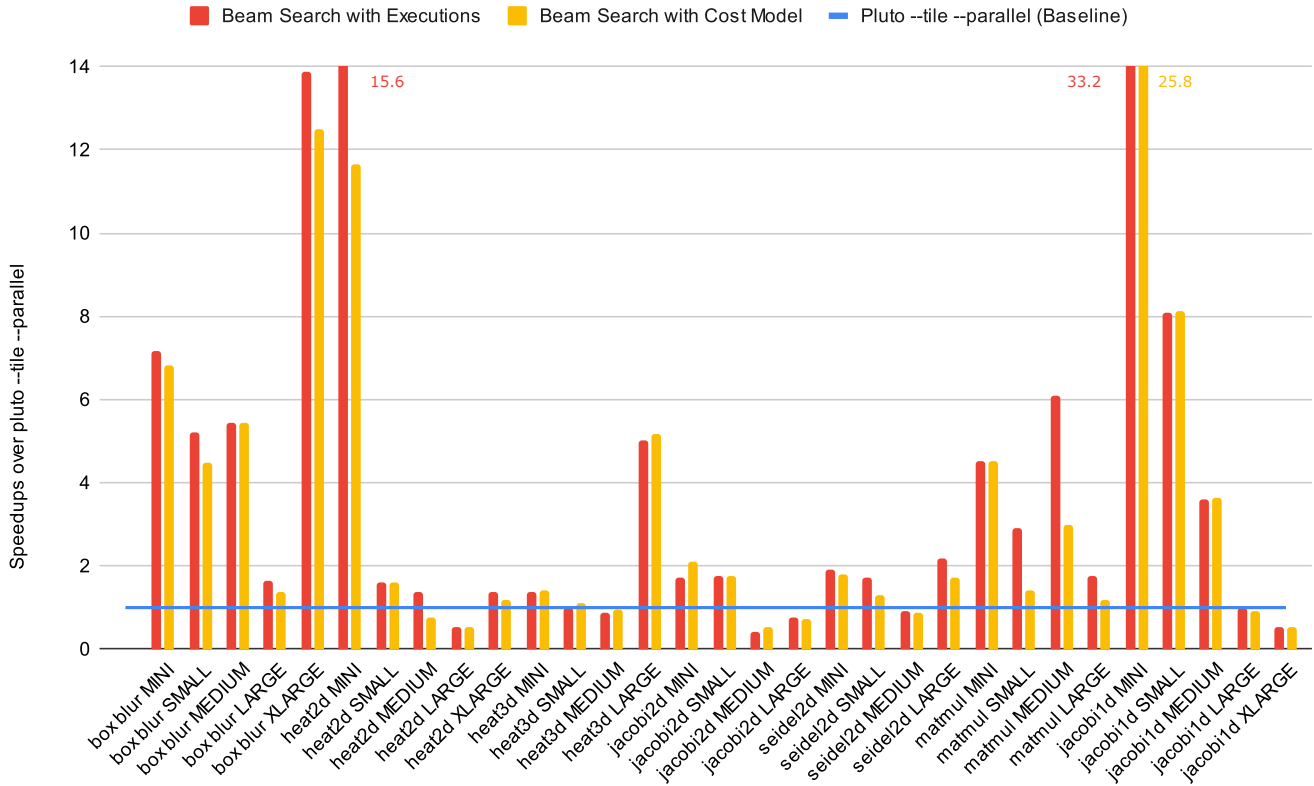

**Figure 1.** Speedups of the schedules found by both the execution guided and cost model guided exploration in comparison with the schedules found by Pluto

based on their effectiveness. For this experiment, we focused on training the model on single computation programs (i.e. loop nests that have a single statement) and evaluating it on single computation benchmarks over different problem sizes. On a test set composed of 2.75 million data points, the proposed model achieves an nDCG score of 0.97.

As shown in figure 1, for most benchmarks our technique achieves higher speedups than Pluto. This is mainly due to the fact that our technique uses a more precise cost model compared to the objective function that Pluto uses (which tries to minimize the distance between producer and consumer statements and maximize outer parallelism).

For example, our cost model takes into consideration the loop nest sizes (extents) when selecting the transformations to apply whereas Pluto does not. For instance, our cost model can detect whether a loop nest is large enough to be parallelized without causing the program to be slowed down because of the overhead of parallelization whereas Pluto always applies parallelization when it is legal. The effects of this difference are considerable in small problem sizes the program should not be parallelized depending on the size of the iteration domain and the amount of work it contains.

Furthermore, being loop size-aware gives our model the ability to detect when it is worth trading-off locality for parallelism and vice versa. This happens in Box Blur XLarge for instance where the extent of the outermost loop is 3 (the number of color channels in the input image). Because Pluto doesn't take into consideration loop extents, it decides to parallelize the outermost loop since it doesn't carry any dependencies. This leads to the program being run on only 3 parallel threads. Whereas our technique chooses to, first, interchange the outermost loop with the subsequent one which has an extent of 2560 (the width of the input image), and then parallelizes the new outermost loop. This makes the program utilize all the available parallel threads on the evaluation machine (which has 48 cores). Although interchanging loops, in this case, deteriorates locality, the efficiency brought by parallelization was worth sacrificing locality as the transformation selected by our technique is, in this example, 14 times faster than the one found by Pluto.

Another key advantage of our technique compared to Pluto is that it can decide whether to apply tiling or not, select the best loops to tile, and choose the best tiling factors depending on the program, whereas in Pluto, a default tiling factor is used across the loop nest whenever possible (if not specified manually). The effects of this difference are noticeable in multiple benchmarks like Heat2D and Matmul.

In some benchmarks, Pluto finds better schedules than the ones found by our technique. This is mainly due to its ability to apply some transformation patterns that we are not considering at this early stage of the project. These unsupported patterns include, for instance, multidimensional skewing (three-dimensional or more) and complex loop reorderings (combination of multiple interchanges). The addition of such patterns to the search space is under-progress and will be included in the next stages of the project.

This experimental evaluation was performed on a multicore dual-socket machine, each socket is a 12-core Intel Xeon E5-2695 v2 CPU with 128 GB RAM total. On this machine, the model-guided exploration is, on average, 252 times faster that the measurements-guided exploration on the proposed set of benchmarks. The smallest improvement in terms of search-speed occurs in small programs where the execution and compilation time of the program is relatively close to inference time of the model. In this experiment, the smallest search-time improvement occurs in Heat3D MINI where model-guided exploration takes 9 seconds whereas the measurements-guided one takes 111 seconds. On the other end, the most important search time improvement is noticeable on large programs and programs with huge search spaces. This occurs in Jacobi2D LARGE for instance where model-guided exploration takes 75 seconds whereas the measurements-guided one takes 63 hours.

Figure 1 also shows that the speedups found by the model-guided exploration are comparable to those found by the measurements-guided exploration for most benchmarks (ground truth, where we execute programs to measure performance instead of using a model). This shows that the model can be used as an accurate surrogate for the execution especially since it provides a very interesting trade-off between search time and the quality of transformations found.

## 5 Conclusion and Future Work

In this paper, we presented a work in progress that consists of expanding the Tiramisu auto-scheduler to support affine loop transformations. Preliminary results of this work show that the proposed approach achieves satisfying results compared to Pluto, a state-of-the-art polyhedral compiler. The next iteration of this project will generalize the current work to cover the whole space of unimodular affine transformations. The next stages also include using more adapted search techniques that, unlike beam search, do not rely upon making local decisions during the exploration.